\documentclass[12pt]{article}
\usepackage{amsmath, amsthm, amssymb}
\usepackage{fullpage}
\usepackage{graphics}
\usepackage{graphicx}
\usepackage{rotating}
\usepackage{mathrsfs}
\usepackage{natbib}
\usepackage{xcolor}
\usepackage{tikz}
\usetikzlibrary{arrows}
\usepackage{wrapfig}
\usepackage{caption}
\usepackage{array}

\renewcommand{\baselinestretch}{1.66}

\newcommand{\logit}{\textrm{logit}}
\def\P{\mathbb{P}}

\def\bagk{\boldsymbol{\alpha}}
\def\vX{\mathbf{X}}

\def\sumin{\sum_{i=1}^n}
\def\expit{\text{expit}}

\newcommand{\be}{\begin{eqnarray}}
\newcommand{\ee}{\end{eqnarray}}

\newcommand{\lily}[1]{\textcolor{blue}{Lily: #1}}

\title{The Impact of Confounder Selection  in Propensity Scores for Rare Events Data - with Applications to Birth Defects}
\author{Ronghui Xu$^{1,2*}$, Jue Hou$^{2}$ and Christina D.~Chambers$^{3,1}$\\
\small  $^1$Department of Family Medicine and Public Health, \\
\small $^2$Department of Mathematics,
\small $^3$Department of Pediatrics,
University of California, San Diego.\\
\small *Correspondence: 9500 Gilman Drive, La Jolla, CA 92093-0112, rxu@ucsd.edu.
}

\begin{document}
   \maketitle

\begin{abstract}

\noindent Our work was motivated by a recent study on birth defects of infants born to  pregnant women exposed to a certain medication for treating chronic diseases. Outcomes such as birth defects are rare events in the general population, which often translate to very small numbers of events in the unexposed group. As drug safety studies in pregnancy are typically observational in nature, we control for confounding in this rare events setting using propensity scores (PS). Using our empirical data, we noticed that the estimated odds ratio for birth defects due to exposure varied drastically depending on the specific approach used. The commonly used approaches with PS are matching, stratification, inverse probability weighting (IPW) and regression adjustment. The extremely rare events setting renders the matching or stratification infeasible. In addition, the PS itself may be formed via different approaches to select confounders from a relatively long list of potential confounders. We carried out simulation experiments to compare different combinations of approaches: IPW or regression adjustment, with 1) including all potential confounders without selection, 2) selection based on univariate association between the candidate variable and the outcome, 3) selection based on change in effects (CIE). The simulation showed that IPW without selection leads to extremely large variances in the estimated odds ratio, which help to explain the empirical data analysis results that we had observed. The simulation also showed that IPW with selection based on univariate association with the outcome is preferred over IPW with CIE. Regression adjustment has small variances of the estimated odds ratio regardless of the selection methods used.

\end{abstract}




\section{Introduction}

Our work was motivated by research  carried out by the Research Center for the Organization of Teratology
Information Specialists (OTIS), which is a North American network of university or hospital
based teratology services that counsel between 70,000 and 100,000 pregnant women every
year.
Research subjects are referred to the Research Center  from the Teratology Information Services and through
other methods of recruitment, where women are consented and  the mothers and their babies are followed  prospectively over time.
Phone interviews are conducted through the length of the pregnancy along with pregnancy
diaries recorded by the mother.
An outcome telephone interview is conducted shortly after the pregnancy ends. If the pregnancy results in a live birth, a dysmorphology exam is performed within the first year of life and with further follow-ups at one year and possibly later dates.

The birth prevalence of major birth defects in the general population is about 3\%, according to
the Centers for Disease Control and Prevention (CDC) Metropolitan Atlanta Congenital Defects Program (MACDP), a population-based birth defects surveillance program and population-based references for secondary endpoints \cite[]{rynn:etal:08}. As pregnant women exposed to a specific medication or other substance in a given recruitment time period are often limited in number, sample sizes in these safety research studies are often limited to as few as 200 subjects in each exposure group, and are powered to detect an odds ratio (OR) of 3 or larger \cite[]{cham:01}. When there is no increased risk of birth defects, this often results in fewer than 10  events in each group.

In a recent study conducted to evaluate the safety in pregnancy for a specific  medication used to  treat certain chronic maternal diseases we had 319  pregnant women who were exposed to the medication and whose pregnancies ended in live birth, and 144 pregnant women who had the  underlying diseases but  were not exposed to the medication and whose pregnancies also ended in live birth. Out of these we had  30 major birth defects in the exposed group, and 5 major birth defects in the unexposed group.

In the cases of observational studies with such rare events, propensity score (PS) methods have been well established in the literature to count for potential confounding \cite[]{brait:rosen:02, cepeda:etal, patorno:etal14}.  These PS based methods generally include matching, stratification, regression adjustment or inverse probability of treatment weighting (IPTW or IPW in general). Due to the extremely rare events in our case, even matching or stratification becomes impractical. IPW on the other hand, has become popular at least partly due to its ease of implementation, since most regression software allow weights as an option. 
In the following we consider regression adjustment and IPW using PS.

Our main concern is to what extent we should perform variable selection in computing the PS. In practice we don't know if an observed variable is truly a confounder \cite[]{GreenlandEtal99}, and different methods have been used to assessing confounding. Two common approaches in practice are: 1) change-in-estimate \cite[CIE]{mick:green:89}, which indirectly assesses the association of the candidate variable with both the exposure and the outcome, since a confounder should be a common cause of both; 2) significance testing of the association between the candidate variable and the outcome only, which was recommended by \cite{Rubin97} in order to reduce the variance of the estimated exposure effect. A third approach is to include all potential confounders. While it has been shown that variables that are only weakly associated with the outcome should not be included in the PS for small studies \cite[]{BrookhartEtal06}, this does not appear to be widely known and confusion persists in practice \cite[]{RotnitzkyEtal10}.

Table \ref{tab:otis} shows the results of analyses using either regression adjustment or IPW with stabilized weights \cite[]{robins:etal:00, hernan:etal:09}, with propensity scores formed by CIE to confirm actual confounders or by simply including all potential confounders collected in the study without any selection or confirmation. The list of all potential confounders is provided in Supplement Table \ref{table:descr_Rsq}. When including all potential confounders due to missing values the sample size is slightly reduced, leading to slightly different crude (i.e. unadjusted) odds ratio (OR) between  exposure to the medication and the outcome of major birth defects.  It is clear that the IPW approach using all potential confounders gives an OR of 6.45 which is very different the other estimated OR's. In the following we carry out simulation experiments to study the behavior of different approaches that are aimed at estimating causal effects of exposure using propensity scores.

\section{Simulation Setup}

Here we restrict our attention to a binary outcome, and the effect measure as commonly used in practice is the  OR.
As logistic regression is commonly used and   will be used to generate data here, we briefly discuss
the non-collapsibility of logistic regression  \cite[]{GreenlandEtal99}. This can be briefly summarized as the discrepancy between the `population averaged' effect and the `conditional' effect under the logistic regression model given other covariates.
Let $A=1$ denote the exposed group, and 0 the unexposed group.
The logistic regression model for the binary outcome $Y$ is
\be\label{eq:logistic}
P(Y=1) = \expit({\alpha}_0+ \alpha_A A + \beta'\vX),
\ee
where $ \expit(x) = e^x/(1+e^x)$ and $\vX$ are the additional covariates.
 The coefficient $\alpha_A$ in the data generating  model, i.e.~the conditional  exposure effect given $\vX$, is often used as the `true' effect in simulation studies for assessing bias and estimation error in general \cite[]{BrookhartEtal06, PirracchioEtal12}. While $\alpha_A$ might be a reasonable target for the regression adjustment approach, we note that it is not the probability limit to which the IPW estimator converges.
In the Appendix we show that IPW estimator converges to the logarithm of the marginal odds ratio between $Y$ and $A$.
This quantity does not generally have closed-form formula based on the logistic regression model (\ref{eq:logistic}) for a given distribution of $\vX$, but can be approximated
 using a very large Monte Carlo sample.

For each simulation scenario below, we will compare the following estimates of the log odds ratio of exposure on outcome: crude, ignoring any covariate information; regression adjustment using PS; IPW using PS; and fitting the multivariate logistic regression model with the true confounders  but without the unobserved $Z$'s (see below). For both regression adjustment and IPW, we consider four different ways of selecting confounders: 1) oracle, i.e.~using the true confounders; 2) CIE, using at least 10\% change as criterion in the estimated OR when adjusting for the potential confounder as compared to the crude OR; 3) significance testing, using $p$-value less than 0.05 as criterion in assess the univariate association of the potential confounder with the outcome in a logistic regression model; 4) including all potential confounders without any selection. The above gives a total of ten different estimators for each simulation scenario.

We consider two types of setups for simulation below: a general one and one based on OTIS data. For each setup, we consider two scenarios: with or without unobserved variables that contribute to the outcome. For the potential confounders, we consider continuous (uniform), binary, as well as categorical distributions.   Notice that categorical variables are associated with multiple coefficients that should be grouped together in any variable selection process \cite[]{YuanLin06, MeierEtal08}. All variables are generated independently.
Table \ref{table:simdata} summarizes the setup of the four scenarios that are detailed in the following. Each scenario was repeated with 10,000 simulation runs.
As scenarios I and II are designed to mimic the rare events structure in the real data, it is not surprising that they are similar to scenario III and IV  in Table \ref{table:simdata}. In general, the average number of total outcome events was between 30 - 35, and well over half of the simulation runs had five or fewer events in the unexposed group. Less than 1\% of the runs had no events in that group.

\subsection{A general setup} 

In scenario I, depicted in Figure \ref{fig:diagram_sim1},  $X_1 \sim$ U(0,1) and $X_2 \sim$ Bernoulli(0.5) are the true confounders. We also generate additional variables $X_3$, $X_4 \sim$ U(0,1), and $X_{5}$ and $X_{6}$ categorical with 3 levels and equal probabilities of $1/3$.
The true propensity model is given by
\be\label{eq:propenI}
\logit P (A=1|\vX) = 1.65  -  1.5X_1 + X_2 - X_3 + 0.6 (X_{5}:2) + 1.2  (X_{5}:3),
\ee
and the true outcome model is
\be\label{eq:outcomeI}
\logit P (Y=1 | A, \vX) = -3.25 + A -  1.5 X_1 - X_2 + 2 X_4 - 0.6 (X_{6}:2) - 1.2  (X_{6}:3),
\ee
where the last two terms in the models above show that level 1 for both $X_{5}$ and $X_{6}$ are used as reference. The coefficients in the models are chosen so that the desired number of events and proportion of exposed subjects are achieved, as summarized in Table \ref{table:simdata}. In addition, we include in the list of potential confounders 7 U(0,1) variables, 10 Bernoulli(0.5) variables, 2 categorical variables with 3 levels, 3 categorical variables with 4 levels, and 2 categorical variables with 5 levels, giving a total of 30 potential confounders for selection purposes. All the categorical variables are with equal probabilities of each level.

From Table \ref{table:simdata} we see that for scenario I there are 31.1 events for the $n=600$ subjects. By design the event rate is lower in the unexposed group. About 0.8\% of the 10,000 simulation runs have generated 0 events in the unexposed group; these are the runs that give estimated odds ratio of infinity. Also about 66\% of the 10,000  runs have generated no more than 5 events in the unexposed group.
The ratio of  exposed versus unexposed numbers of subjects is about 2:1 as in the real data.

Next we add unobserved variables for scenario II, as depicted in Figure \ref{fig:diagram_sim2}.
The setup is otherwise the same as in scenario I, with the same regression coefficients as in (\ref{eq:propenI}) and (\ref{eq:outcomeI}), but with the addition of $Z_1$, $Z_2$ and $Z_3 \sim$ N(0, 0.25), each with a coefficient of one.
From Table \ref{table:simdata} we see that the distribution of number of events is similar to scenario I.

\subsection{Setup based on OTIS data}

\begin{figure}[htb]
\begin{center}
\begin{tikzpicture}[
  font=\sffamily\footnotesize,
  every matrix/.style={ampersand replacement=\&,column sep=-0.3cm,row sep=0.5cm},
  source/.style={draw,thick,rounded corners, fill=white,inner sep=.2cm},
  process/.style={draw,thick,circle,fill=white,inner sep=.1cm},
  sink/.style={source,fill=green!20},
  datastore/.style={draw,very thick,shape=datastore,inner sep=.3cm},
  dots/.style={gray,scale=2},
  to/.style={->,>=stealth',shorten >=1pt,semithick,font=\sffamily\footnotesize},
  every node/.style={align=center}]
  \matrix{
     \node[source] (A1) {Referral Source};
      \&\&  \node[source] (A2) {Asthma \\ Maternal Height};
      \&\&\\
 	 \& \node[process] (X) {Exposure};
	\&\& \node[process] (Y) {Outcome}; \\
  };
  	\draw[to] (A2) -- (X);
	\draw[to] (A2) -- (Y);
  	\draw[to] (A1) -- (X);
	\draw[to] (X) -- (Y);
\end{tikzpicture}
\end{center}
\caption{Diagram for Scenario III}\label{fig:diagram_OTIS1}
\end{figure}
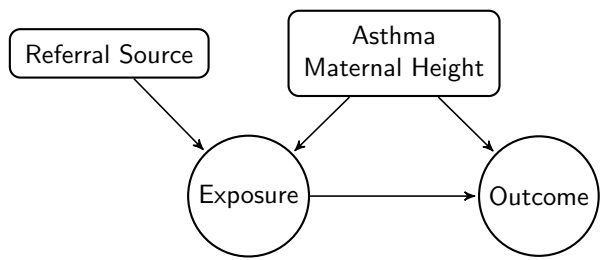

In this setup we consider the 37 potential confounders from the OTIS study that are given in the Supplement Table \ref{table:descr_Rsq} for $n=439$ subjects. Based on the final analysis of the original dataset, asthma and maternal height were selected using CIE as the confirmed confounders. In addition, referral source was found to be relatively strongly correlated with exposure \cite[generalized $R^2 = 16.2\%$ in Supplement Table \ref{table:descr_Rsq}]{cox:snell}. When the models based on Figure \ref{fig:diagram_OTIS1} are fitted to the original data, we have
\be\label{eq:OTISpropen}
\logit P (A=1|\vX) &=& 9.68  - 0.54 * \mbox{asthma} - 0.05 * \mbox{mat\_height} \\
& - & 0.45 * \mbox{ref}_1 + 0.34 * \mbox{ref}_2 - 17.19 * \mbox{ref}_3 + 2.14 * \mbox{ref}_4 - 1.93 * \mbox{ref}_5, \nonumber
\ee
where the six levels of referral sources are: 0 - health-care professional, 1 - internet, 2 - other, 3 - patient support group, 4 - pharmaceutical company / sponsor, and 5 - TIS;
and
\begin{equation}\label{eq:outcome_dep on other}
\logit P (Y=1 | A, \vX) = 8.23 + 1.03 A + 1.59 * \mbox{asthma} - 0.07 * \mbox{mat\_height}.
\end{equation}
Models (\ref{eq:OTISpropen}) and (\ref{eq:outcome_dep on other}) are then used in scenario III as the true propensity and the true outcome model to generate simulated data.
The number of events under this scenario is again summarized in Table \ref{table:simdata}.

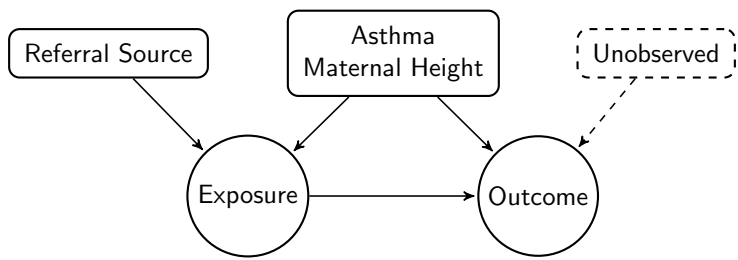
\begin{figure}[htb]
\begin{center}
\begin{tikzpicture}[
  font=\sffamily\footnotesize,
  every matrix/.style={ampersand replacement=\&,column sep=-0.3cm,row sep=0.5cm},
  source/.style={draw,thick,rounded corners, fill=white,inner sep=.2cm},
  unknown/.style={draw,thick, dashed,rounded corners, fill=white,inner sep=.2cm},
  process/.style={draw,thick,circle,fill=white,inner sep=.1cm},
  sink/.style={source,fill=green!20},
  datastore/.style={draw,very thick,shape=datastore,inner sep=.3cm},
  dots/.style={gray,scale=2},
  to/.style={->,>=stealth',shorten >=1pt,semithick,font=\sffamily\footnotesize},
  every node/.style={align=center}]
  \matrix{
     \node[source] (A1) {Referral Source};
      \&\&  \node[source] (A2) {Asthma \\ Maternal Height};
      \&\&  \node[unknown] (A3) {Unobserved}; \\
 	 \& \node[process] (X) {Exposure};
	\&\& \node[process] (Y) {Outcome}; \\
  };
  	\draw[to] (A2) -- (X);
	\draw[to] (A2) -- (Y);
  	\draw[to] (A1) -- (X);
  	\draw[dashed,to] (A3) -- (Y);
	\draw[to] (X) -- (Y);
\end{tikzpicture}
\end{center}
\caption{Diagram for Scenario IV}\label{fig:diagram_OTISglmm}
\end{figure}

Finally for scenario IV we consider an unobserved variable that contributes to the outcome. This is motivated by the fact that the generalized $R^2=0.16$ for the above outcome model (\ref{eq:outcome_dep on other}),
indicating that about 84\% of the variation in the outcome remains unexplained.
Fitting a logistic regression model with a normally distributed random intercept $Z$ to account for the unobserved heterogeneity, we have
\be\label{eq:OTIS_unobs}
\logit \P (Y=1 | A, \vX) = 10.5  + 1.48 A + 2.29 * \mbox{asthma} - 0.10 * \mbox{mat\_height} + Z ,
\ee
where $Z \sim$ N(0, 3.9).  We note that the estimated variance of $Z$ in this case is quite large, and the estimated exposure effect has increased substantially from 1.03 (SE = 0.514) to 1.48 (SE = 2.569). Although the estimated exposure effect is   no longer significantly different from zero,
 we still use  model (\ref{eq:OTIS_unobs}) with the fitted point estimates as the true outcome model to generate data as depicted in Figure \ref{fig:diagram_OTISglmm}.
Despite the very different coefficients in the outcome models (\ref{eq:outcome_dep on other}) and (\ref{eq:OTIS_unobs}), Table \ref{table:simdata} once again shows similar numbers of outcome events as for the previous scenarios, perhaps reflecting the fact that the same original data was used to create both outcome models.

\section{Simulation Results}

Table \ref{table:selection}  summarizes the results of confounder selection, using both CIE and significance testing. The CIE selected roughly $2$ potential confounders on average (i.e.~true positive $+$ false positive). The univariate $p$-value selected a few more, between $3$ and $4$.
In terms of accuracy, both methods had a reasonable chance, about $50\%$ or more, to include all true confounders in their selection (column `Incl.').
However, they tended to choose non-confounding variables as well.
Their chances of selecting exactly the set of true confounders were below $30\%$  (column `Exact').
As expected the univariate p-value rule had lower rate of exact capture since it had larger average number of false positives.

Table \ref{table:top5} lists the top five selected confounders for each scenario and each selection method, i.e.~CIE or PVAL (for univariate $p$-value).
Compared to the diagrams of each scenario, it is clear  that in addition to the true confounders,  CIE had a tendency to selection the `instrumental variables' (variables that affect the outcome only through their effects on exposure): $X_3$, $X_5$ in scenarios I and II, and referral source in scenarios III and IV. In contrast, significance testing tended to selection those variables that contributed to the outcome, even though they were not associated with exposure ($X_4$ and $X_6$).

\begin{figure}[htp]
  \centering
  \includegraphics[width=1.0\textwidth]{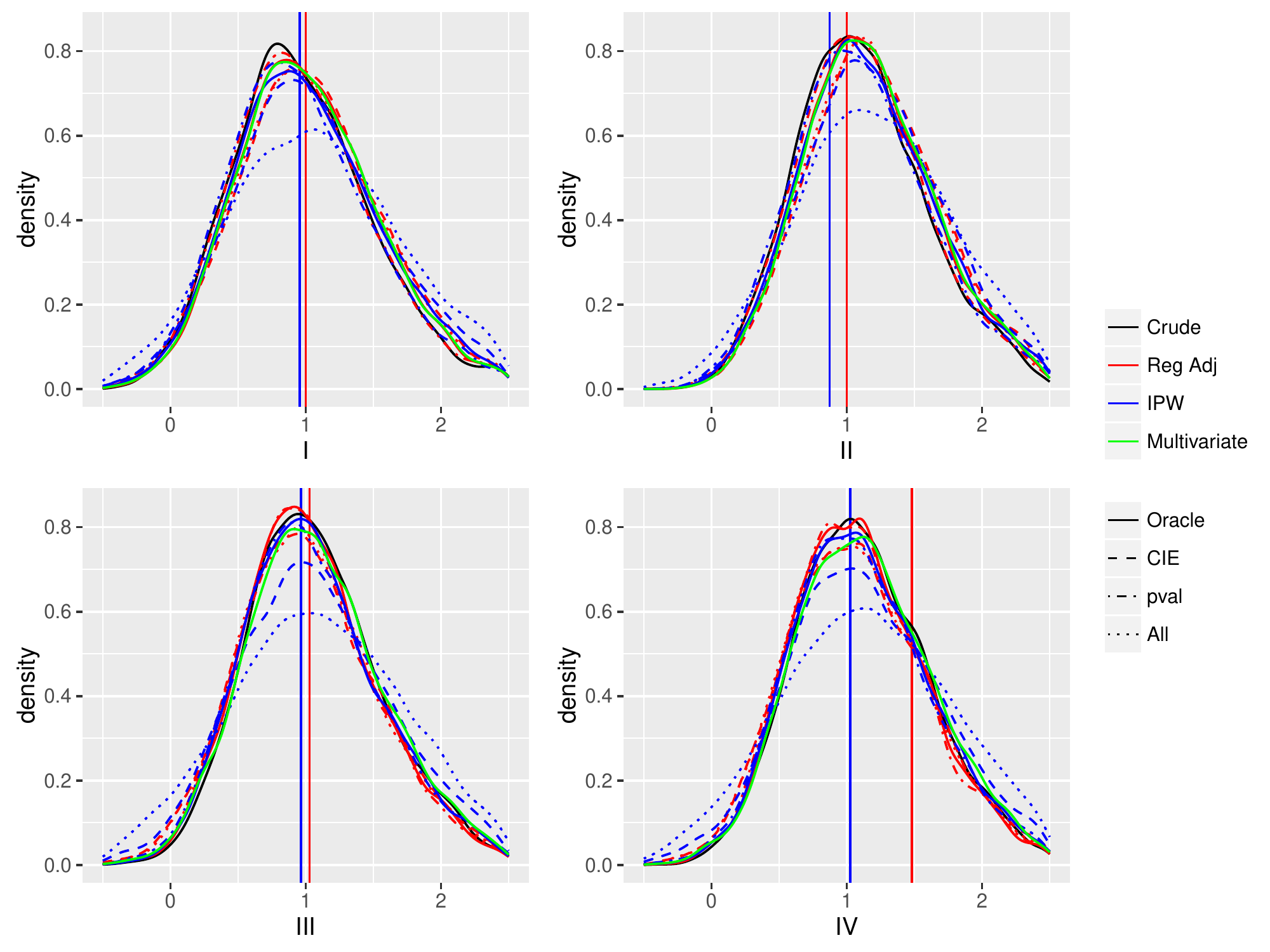}
  \caption{Kernel smoothed density of the estimates from simulation;
  the vertical lines are the true conditional (red) and marginal (blue) effects, respectively. The line types in the legend refer to red and blue only. }\label{fig:dens}
\end{figure}

\begin{figure}[htp]
  \centering
  \includegraphics[width=1.0\textwidth]{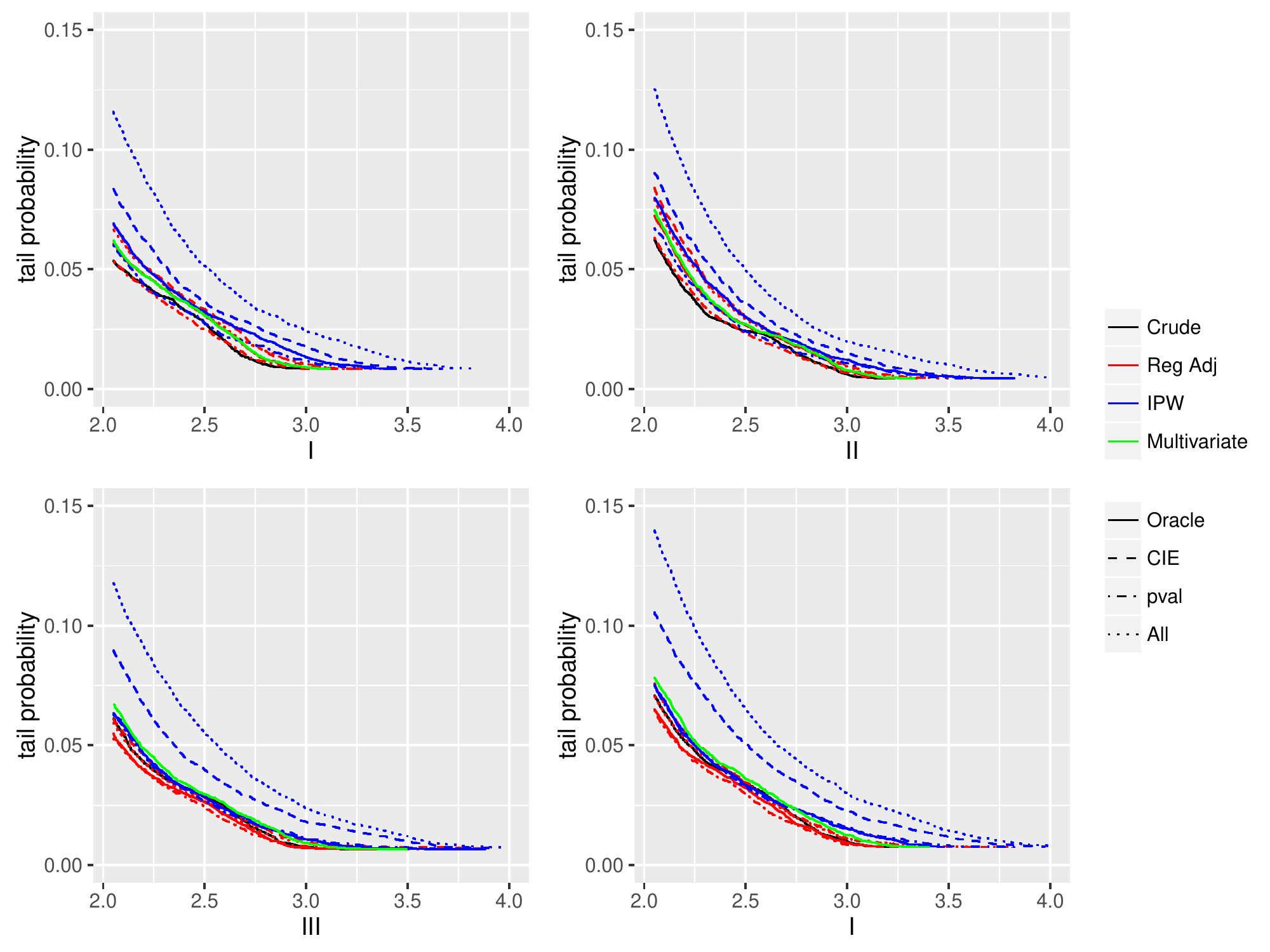}
  \caption{Tail probabilities of the estimates from simulation.
  The line types in the legend refer to red and blue only.   }\label{fig:tail}
\end{figure}

Figure \ref{fig:dens} shows the distribution of the ten different estimators described earlier for each scenario. Common to all four scenarios is that the IPW approach using all potential confounders without any selection  had the largest spread among the ten, followed by IPW using CIE to selection confounders.
This is also confirmed in Figure \ref{fig:tail} for the tail probabilities, i.e. one minus the empirical cumulative distribution function. Note that the tail probabilities flattened out at the frequency of simulation runs with no events in the unexposed group (Table \ref{table:simdata}), in which case all ten estimates were infinite.

As discussed earlier, even with the same estimated PS, the regression adjustment and the IPW approach estimate different quantities,  one conditional and one marginal (vertical lines in Figure \ref{fig:dens}). It is known that when the conditional logistic model is true - which is the case by design of the simulations - the marginal effect ignoring covariates is typically biased towards zero \cite[]{robinson:jewell}. This discrepancy is seen to be particularly outstanding in scenario IV, and it is interesting to observe that when ignoring the unobserved $Z$ in the PS,
even the regression adjustment estimates seem to be centered closer to the marginal effect.

\section{Discussion}

A confounder is a covariate that affects the quantity of interest such as a population mean of the outcome, that differs between the exposed (i.e treated) and unexposed populations, and that this difference between the populations has led to confounding of an association measure for the effect of interest \cite[]{GreenlandEtal99}. As we stated earlier, whether a potential confounder is a true confounder is unknown in practice. Causal knowledge should be a prerequisite for confounder assessment \cite[]{hernan:etal:02}, and is used to create our list of potential confounders. Given this list, additional criteria are needed to assess the more mathematical aspects of confounding.
 In our medication and vaccine safety studies,  CIE together with the assessment of  correlation between a potential confounder and both the exposure and the outcome variables are routinely used to identity confounders. Indeed CIE alone does not imply confounding, especially without causal knowledge. Another concern about CIE is the non-collapsibility discussed earlier. However, in our experience the change in estimates due to  non-collapsibility tend to be under 10\% which is our criterion cutoff; this was also confirmed in our simulations when there were no unobserved confounders, so that the marginal and the conditional effects are sufficiently close \cite[]{PirracchioEtal12}.

 Our simulation results clearly show that  the IPW approach using all potential confounders without any selection has the greatest variability. This should help to explain the extremely large estimated OR of 6.45 observed in Table \ref{tab:otis}. In addition, and
in particular in scenarios III and IV, we see that if the IPW is used, then the univariate assessment of the correlation between a potential confounder and the outcome is preferred over the CIE, at least in the rare events situations considered in this paper. The simulation results also shows that CIE has some tendency to select what is referred to as instrumental variables in the literature, i.e.~variables that affect the outcome only through their effects on the exposure, which are known not to be included in the propensity scores. Finally, regression adjustment appears to have small variances of the estimated odds ratio regardless of the selection methods used.

The weighted approach was initially proposed in \cite{horv:thom:52} and has continued to be studied in the survey research literature \cite[]{gelman2007}. As \cite{kang:schaf} pointed out, surveys are usually designed to ensure that IPW estimates are acceptably precise, but in the more general missing data problems it has been known since at least the 1980s that IPW methods can assign relatively large weights to  certain observations leading to large variances of the effect estimates.
In our case it was those 5 unexposed subjects who had a major birth defect outcome that received lower than usual (stabilized) weights.
As we have illustrated here, even using the stabilized weights does not solve the large variance problem. For the IPW approach using PS,
\cite{RotnitzkyEtal10} showed asymptotically that adjusting for a covariate is efficient if the covariate is independent of the exposure, while
not adjusting is efficient if the covariate is independent of the outcome given the exposure level.
Similar conclusions have also been reached via empirical investigations \cite[]{BrookhartEtal06}.

A main concern for the regression adjustment  approach is that the regression model for the outcome might be wrong; however, \cite{vansteel14} showed that the standard test of the null hypothesis of no exposure effect (using robust variance estimators), as well as particular standardized effects obtained from such adjusted regression models, are robust against misspecification of the outcome model as long as the PS model is correctly specified. We note that the correct PS model is required for all PS-based methods to be valid. For rare events like in our settings, \cite{xu:etal14} recommended the regression adjustment  approach.



For outcomes with typically rare events in the population, such as major birth defects, the numbers of events in the unexposed groups are likely very small, as seen in the drug safety study that motivated this paper. In the future we might consider approaches to increase the sample sizes of the
unexposed groups. This, however, might be limited by the feasibility to  recruit  in a given time period pregnant women with a certain disease and without exposure to the medication under study \cite[]{cham:etal:10}.  The inclusion of   historical controls, on the other hand, might bring in additional confounding that needs to be controlled for.


\vskip .3in
\centerline{\bf Acknowledgement}
We appreciate our discussion with the US Food and Drug Administration (FDA) statisticians regarding the drug safety study that motivated this work, and their encouragement for us to publish the research results.
\vskip .1in

\vskip .3in
\centerline{\large APPENDIX}
\vskip .1in

 Write $w_i$ as the weight for the $i$-th subject.
Straightforward algebra shows that the weighted score equations which the IPW estimator $\hat{\bagk} = (\hat{\alpha}_0, \hat{\alpha}_A)^\top$
  solves can be written as
  $$
  \left\{
  \begin{array}{>{\displaystyle}c >{\displaystyle}l }
   \sumin \{Y_i - \expit(\hat{\alpha}_0+\hat{\alpha}_A)\} A_i w_i &= 0 , \\
   \sumin \{Y_i - \expit(\hat{\alpha}_0)\} (1-A_i) w_i &= 0.
  \end{array}
\right.
$$
Therefore
$$
\hat{\alpha_A} = \logit\left(\frac{n^{-1}\sumin Y_i A_i w_i}{n^{-1}\sumin A_i w_i}\right)
-  \logit\left\{\frac{n^{-1}\sumin Y_i (1-A_i) w_i}{n^{-1}\sumin (1-A_i) w_i}\right\}.
$$
Using stabilized weights
$$
w_i = \frac{A_i \widehat{P}(A=1) }{\hat\pi_i} + \frac{ (1-A_i) \widehat{P}(A=0) }{1-\hat\pi_i},
$$
where $\pi_i = P( A=1 | \vX_i)$ is specified under the propensity score model, and $ \widehat{P}(A=1)$ is the empirically estimated proportion of $A=1$. Assume  correct specification of the PS model, it can be seen that
the IPW estimator
 converges to the following population averaged quantity:
$$ 
\logit \{ E_{\vX}[P(Y=1 | A=1, \vX)] \}  - \logit \{ E_{\vX}[P(Y=1 | A=0, \vX)] \},
$$ 
which is the logarithm of the marginal OR between $Y$ and $A$.

\bibliographystyle{natbib}
\bibliography{propensity}

\newpage
\renewcommand{\baselinestretch}{1.0}

\begin{table}
  \caption{Estimated Odds Ratio (95 \% CI) of Birth Defects  with Different Approaches Using Propensity Scores }\label{tab:otis}
$$\begin{tabular} {lccc}
\hline\hline\\
   &Crude & Reg.~Adjustment & IPW \\
\hline\\
Selection by CIE$^1$ &  2.89 & 2.77 & 3.38 \\
  & (1.10, 7.60) & (1.08, 7.13) & (1.27, 9.00)\\
\hline\\
All Potential Confounders$^2$ & 2.87 & 3.00 & {6.45} \\
 & (1.09, 7.58) & (1.22, 7.39) & (2.26, 18.36) \\
\hline
{\footnotesize $^1$change-in-effect, see text for more description}\\
{\footnotesize $^2$given in Supplement Table \ref{table:descr_Rsq} }
\end{tabular}$$
\end{table}

\begin{figure}[htb]
\begin{center}
\begin{tikzpicture}[
  font=\sffamily\footnotesize,
  every matrix/.style={ampersand replacement=\&,column sep=-0.3cm,row sep=0.5cm},
  source/.style={draw,thick,rounded corners, fill=white,inner sep=.5cm},
  process/.style={draw,thick,circle,fill=white,inner sep=.5cm},
  sink/.style={source,fill=green!20},
  datastore/.style={draw,very thick,shape=datastore,inner sep=.3cm},
  dots/.style={gray,scale=2},
  to/.style={->,>=stealth',shorten >=1pt,semithick,font=\sffamily\footnotesize},
  every node/.style={align=center}]
  \matrix{
     \node[source] (A1) {$X_3$,  $X_{5}$};
      \&\&  \node[source] (A2) {$X_1$,  $X_2$};
      \&\&  \node[source] (A3) {$X_4$,  $X_{6}$}; \\
 	 \& \node[process] (X) {A};
	\&\& \node[process] (Y) {Y}; \\
  };
  	\draw[to] (A2) -- (X);
	\draw[to] (A2) -- (Y);
  	\draw[to] (A1) -- (X);
  	\draw[to] (A3) -- (Y);
	\draw[to] (X) -- (Y);
\end{tikzpicture}
\end{center}
\caption{Diagram  for Scenario I}\label{fig:diagram_sim1}
\end{figure}
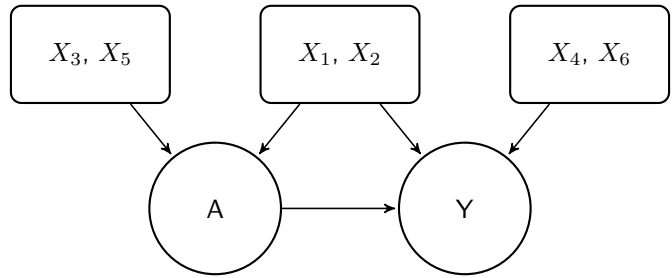

\begin{figure}[htb]
\begin{center}
\begin{tikzpicture}[
  font=\sffamily\footnotesize,
  every matrix/.style={ampersand replacement=\&,column sep=-0.3cm,row sep=0.5cm},
  source/.style={draw,thick,rounded corners, fill=white,inner sep=.5cm},
  unknown/.style={draw,thick, dashed,rounded corners, fill=white,inner sep=.5cm},
  process/.style={draw,thick,circle,fill=white,inner sep=.5cm},
  sink/.style={source,fill=green!20},
  datastore/.style={draw,very thick,shape=datastore,inner sep=.3cm},
  dots/.style={gray,scale=2},
  to/.style={->,>=stealth',shorten >=1pt,semithick,font=\sffamily\footnotesize},
  every node/.style={align=center}]
  \matrix{
 \node[source] (A1) {$X_3$,  $X_{5}$};
      \&\&  \node[source] (A2) {$X_1$,  $X_2$};
      \&\&  \node[source] (A3) {$X_4$,  $X_{6}$} ;  \\
 \& \node[process] (X) {A};
	\&\& \node[process] (Y) {Y}; \&
	\\
\node[unknown](Z1){$Z_1$};\&  \& \node[unknown](Z2){$Z_2$};
  \&  \& \node[unknown](Z3){$Z_3$}; \\
  };
  	\draw[to] (A2) -- (X);
	\draw[to] (A2) -- (Y);
  	\draw[to] (A1) -- (X);
  	\draw[to] (A3) -- (Y);
	\draw[to] (X) -- (Y);
	\draw[dashed,to] (Z1) -- (X);
	\draw[dashed,to] (Z2) -- (X);
	\draw[dashed,to] (Z2) -- (Y);
	\draw[dashed,to] (Z3) -- (Y);
\end{tikzpicture}
\end{center}
\caption{Diagram  for Scenario II}\label{fig:diagram_sim2}
\end{figure}
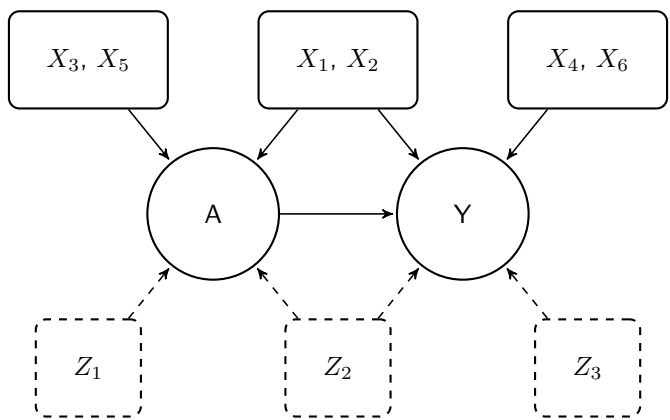

\begin{table}[htb]
\caption{Summary of Simulation Setup }\label{table:simdata}
\begin{center}
{\small
\begin{tabular}{ccc|ccc}
\hline\hline
\multicolumn{3}{c|}{ } & \multicolumn{3}{c}{Number of  events }\\
 Scenario & $n$ & $p^*$ & Average (SD) &  None in Unexposed &  $\le$ 5 in Unexposed\\
\hline
I & 600 & 30 & 31.1 (5.5) & 0.8\% & 66\%\\
II & 600 & 30 & 30.8 (5.4) & 0.7\% & 63\%\\
III & 439 & 37 & 34.1 (5.5) & 0.6\% & 59\%\\
IV & 439 & 37 & 34.0 (5.4) & 0.7\% & 63\%\\
\hline
\end{tabular}
}
\end{center}
$^*$ $p$ is the number of potential confounders under consideration. 
\end{table}

\begin{table}[htb]
\caption{Summary of Confounder Selection}\label{table:selection}
\begin{center}
{\small
\begin{tabular}{ccccccccc}
\hline\hline
& \multicolumn{4}{c}{CIE} & \multicolumn{4}{c}{Univariate $p$-value}\\
 Scenario   & True Pos. & False Pos.  & Incl. & Exact &
 True Pos. & False Pos. & Incl. & Exact \\
\hline
 I & 1.60 & 0.69 & 63\% & 30\% & 1.30 & 2.84 & 42\% & 0.7\% \\
  II & 1.53 & 0.63 & 57\% & 28\% & 1.27 & 2.80 & 41\% & 0.8\% \\
  III & 1.55 & 0.88 & 60\% & 18\% & 1.83 & 1.94 & 84\% & 12.5\% \\
  IV & 1.54 & 0.90 & 59\% & 17\% & 1.82 & 1.91 & 82\% & 13.0\% \\
\hline
\end{tabular}
}
\end{center}
\end{table}

\begin{table}[htb]
\caption{Frequencies of Top Five Selected Confounders}\label{table:top5}
\begin{center}
{\small
\begin{tabular}{ccccccc}
\hline\hline
 Scenario   & Method & 1 &   2 & 3 & 4 & 5 \\
\hline
I & CIE & $X_2$ & $X_1$ & $X_3$ & $X_5$ & $X_4$\\
&& 89.35 \% & 69.19 \% & 28.27 \% & 22.31 \% & 7.35 \%\\
 & PVAL & $X_4$ & $X_1$ & $X_6$ & $X_2$ & $X_3$\\
&& 81.96 \% & 70.88 \% & 59.35 \% & 58.15 \% & 7.61 \%\\
\\
II & CIE & $X_2$ & $X_1$ & $X_3$ & $X_5$ & $X_4$\\\
&& 87.85 \% & 64.18 \% & 24.81 \% & 18.59 \% & 7.35 \%\\
 & PVAL & $X_4$ & $X_1$ & $X_2$ & $X_6$ & $X_3$\\
&& 81.40 \% & 69.53 \% & 59.26 \% & 58.49 \% & 7.14 \%\\
\\
III & CIE & Asthma & Mat\_height & Referral & State & Education \\
&& 80.11 \% & 73.15 \% & 62.34 \% & 3.56 \% & 2.82 \%\\
 & PVAL & Asthma & Mat\_height & SES & IVF & Mat\_weight\\
&& 93.38 \% & 87.49 \% & 11.87 \% & 10.81 \% & 9.61 \%\\
\\
IV & CIE & Asthma & Mat\_height & Referral & State & Education \\
&& 80.87 \% & 70.95 \% & 61.78 \% & 3.63 \% & 3.06 \%\\
 & PVAL & Asthma & Mat\_height & SES & IVF & Referral\\
&& 93.65 \% & 86.05 \% & 11.43 \% & 10.42 \% & 10.11 \%\\
\hline
\end{tabular}
}
\end{center}
\end{table}


\newpage

\begin{table}
\caption{Supplement - List of All Potential Confounders and Their Association with Exposure and Outcome}\label{table:descr_Rsq}
{\footnotesize
$$\begin{tabular} {lc|cc}
\hline
\hline
 \multicolumn{2}{c|}{} & \multicolumn{2}{c}{Generalized $R^2$ with}\\
Variable & \# Levels$^1$ & Exposure& Outcome\\
\hline
 Asthma & 2 & 0.6 \% & 2.8 \% \\
  Maternal Height & -- & 2.1 \% & 2.3 \% \\
  Maternal Age Group & 4 & 0.8 \% & 2.3 \% \\
  Race & 4 & 0.4 \% & 1.9 \% \\
  Co-Medication Duration in 1st Trimester & 5 & 0.4 \% & 1.7 \% \\
  Education & 3 & 0.4 \% & 1.3 \% \\
  Multiple Birth & 2 & 0.6 \% & 1.1 \% \\
  Years Since Diagnosis of Primary Disease & -- & 0 \% & 1 \% \\
  Disease Severity Score 1 at Intake & -- & 0.2 \% & 0.9 \% \\
  Disease Severity Score 2 at Intake & -- & 0 \% & 0.8 \% \\
  Referral Source & 7 & 16.2 \% & 0.8 \% \\
  IVF & 2 & 0.6 \% & 0.7 \% \\
  History of Birth Defects & 2 & 0 \% & 0.7 \% \\
  Smoking in 1st Trimester & 2 & 0 \% & 0.7 \% \\
  Infections in 1st Trimester & 3 & 0.2 \% & 0.7 \% \\
  Disease Severity Score 3 at Intake & -- & 0 \% & 0.7 \% \\
  Disease Severity Score 4 at Intake & -- & 0.1 \% & 0.4 \% \\
  SES & 3 & 0.2 \% & 0.3 \% \\
  Number of Other Diseases Diagnosed & 2 & 0 \% & 0.3 \% \\
  Vitamin & 3 & 1.1 \% & 0.3 \% \\
  Thyroid Disease & 2 & 0.1 \% & 0.3 \% \\
  Psychiatric Conditions & 2 & 0 \% & 0.2 \% \\
  Gestational Age at Enrollment & -- & 0 \% & 0.2 \% \\
  Intended Pregnancy & 2 & 0.3 \% & 0.2 \% \\
  Co-Medication Average Dose in 1st Trimester & -- & 0.1 \% & 0.1 \% \\
  BMI & 4 & 1.3 \% & 0.1 \% \\
  Pre-gestational Hypertension & 2 & 0 \% & 0.1 \% \\
  Previous Spontaneous Abortion & 2 & 0 \% & 0.1 \% \\
  Other Specific Diseases & 2 & 0.2 \% & 0.1 \% \\
  Amniocentesis & 2 & 0.7 \% & 0.1 \% \\
  Chorionic Villus & 2 & 0.3 \% & 0.1 \% \\
  Primary Disease & 2 & 2.2 \% & 0 \% \\
  Alcohol in 1st Trimester & 2 & 0 \% & 0 \% \\
  Other Major Teratogens in 1st Trimester & 2 & 0.6 \% & 0 \% \\
  Gender of Birth & 3 & 0.5 \% & 0 \% \\
  Ultrasound Level 2 & 2 & 0.3 \% & 0 \% \\
  Gravidity & 2 & 0 \% & 0 \% \\
  Disease Severity Score 5 at Intake & -- & 0.3 \% & 0 \% \\
  Disease 1 Severity Score Imputation Indicator & 2 & 2.8 \% & 0 \% \\
  Year of Enrollment & -- & 0.1 \% & 0 \% \\
  Disease 2 Severity Score Imputation Indicator & 2 & 0.3 \% & 0 \% \\
  Country & 3 & 1.1 \% & 0 \% \\
  Parity & 2 & 0 \% & 0 \% \\
  Maternal Weight & -- & 0.2 \% & 0 \% \\
\hline
\end{tabular}$$
$^1$ `--' for continuous variables.
}
\end{table}

\end{document}